\documentclass[english,aps,floatfix]{revtex4}
\usepackage[T1]{fontenc}
\usepackage[latin9]{inputenc}
\usepackage{graphicx}

\usepackage{amsfonts}

\usepackage[dvips]{epsfig}

\usepackage{babel}

\begin{document}

\title{Quantum dice rolling: A multi-outcome generalization of quantum coin
flipping}

\author{N. Aharon}

\author{J. Silman}

\affiliation{School of Physics and Astronomy, Tel-Aviv University, Tel-Aviv 69978,
Israel}
\begin{abstract}
\textbf{We generalize the problem of coin flipping to more than two
outcomes and parties. We term this problem dice rolling, and study
both its weak and strong variants. We prove by construction that in
quantum settings (i) weak $N$-sided dice rolling admits an arbitrarily
small bias for any value of $N$, and (ii) two-party strong $N$-sided
dice rolling saturates the corresponding generalization of Kitaev's
bound for any value of $N$. In addition, we make use of this last
result to introduce a family of optimal $2m$-party strong $n^{m}$-sided
dice rolling protocols for any value of $m$ and $n$.}
\end{abstract}
\maketitle

\section{Introduction}

Coin flipping (CF) is a cryptographic problem in which a pair of remote
distrustful parties, usually referred to as Alice and Bob, must generate
a random bit that they agree on. There are two types of coin flipping
protocols. In weak CF one of the parties prefers one of the outcomes
and the other prefers the opposite, whereas in strong CF each party
does not know the other's preference. The security of a CF protocol
is quantified by the biases $\epsilon_{A}^{\left(i\right)}$ and $\epsilon_{B}^{\left(i\right)}$
($i=0,\,1$); if $\left.P_{A}^{\left(i\right)}\right.^{*}$ and $\left.P_{B}^{\left(i\right)}\right.^{*}$
are the maximal probabilities that a dishonest Alice or Bob can force
the outcome $i$, then \begin{equation}
\epsilon_{j}^{\left(i\right)}\hat{=}\left.P_{j}^{\left(i\right)}\right.^{*}-\frac{1}{2}\,,\qquad i\in\left\{ 0,\,1\right\} \,.\end{equation}
 The biases tell us to what extent each of the parties can increase
beyond one half their chances of obtaining each of the outcomes. In
weak CF we associate each of the outcomes with a win of one party
over the other. Hence, we are not interested in bounding the maximal
losing probability and we consequently consider only two biases: $\epsilon_{i}\hat{=}P_{i}^{*}-1/2$,
where $P_{i}^{*}$ is dishonest party $i$'s maximal winning probability.
$\epsilon\hat{=}\max\left(\epsilon_{A},\,\epsilon_{B}\right)$, or
$\epsilon\hat{=}\max\bigl(\epsilon_{A}^{\left(0\right)},\,\epsilon_{A}^{\left(1\right)},\,\epsilon_{B}^{\left(0\right)},\,\epsilon_{B}^{\left(1\right)}\bigr)$
in the case of strong CF, is often referred to as the bias of the
protocol.\\

The problem of CF was first introduced by Blum in 1981, who analyzed
it in classical settings \cite{Blum}. It was subsequently shown that
if there are no limitations on the parties' computational power a
dishonest party can always force any outcome they desire \cite{Kilian}.
With the publication of the quantum key distribution protocol of Bennett
and Brassard in 1984 \cite{BB84}, it was realized that many communication
tasks that are impossible in a classical setting may be possible in
a quantum setting. In 1999 Goldenberg \emph{et al.} introduced a quantum
gambling protocol \cite{Gambling}, which is a problem closely related
to weak CF (see \cite{footnote}). The first quantum (strong) coin
flipping protocol per se was presented by Aharonov \emph{et al.} in
2000 \cite{Aharonov}. The protocol achieves a bias of $\sqrt{2}/4\simeq0.354$
\cite{Spekkens -2}. Soon afterward Spekkens and Rudolph \cite{Spekkens -1},
and independently Ambainis \cite{Ambainis}, devised a strong CF protocol
with a bias of $1/4$. On the other hand, Kitaev subsequently proved
that there is a limit to the efficacy of strong CF protocols \cite{Kitaev}:
Any strong CF protocol must satisfy $\left.P_{A}^{\left(i\right)}\right.^{*}\cdot\left.P_{B}^{\left(i\right)}\right.^{*}\geq1/2$,
$i\in\left\{ 0,\,1\right\} $. As regards weak CF, in 2002 Spekkens
and Rudolph introduced a family of three rounds of communication protocols
in which both dishonest parties have a bias of $(\sqrt{2}-1)/2\simeq0.207$
\cite{Spekkens}. Mochon then improved upon Spekkens and Rudolph's
result by constructing weak CF protocols with an infinite number of
rounds \cite{Mochon_-2,Mochon_-1}. These efforts culminated in a
proof that weak CF with an arbitrarily small bias is possible \cite{Mochon}.
Most recently, building upon Mochon's latest result, Chailloux and
Kerenidis devised a strong CF protocol, which saturates Kitaev's bound
in the limit of an infinite number of rounds \cite{Kerenidis}.\\

CF admits many generalizations. We may consider imbalanced coins,
more than two outcomes, more than two parties, or indeed some or all
of these possibilities combined. An analysis of multi-party strong
CF -- the problem of $N$ remote distrustful parties having to decide
on a bit -- in quantum settings was carried out in \cite{Multi-party},
and similarly to the two-party case, it was shown that the use of
quantum resources is advantageous. Even though it is usually mentioned
only with respect to the balanced case, the general statement of Kitaev's
bound reads \begin{equation}
\left.P_{A}^{\left(i\right)}\right.^{*}\cdot\left.P_{B}^{\left(i\right)}\right.^{*}\geq P_{i}\,,\qquad i\in\left\{ 0,\,1\right\} \,,\end{equation}
 where $P_{i}$ is the probability of the outcome $i$ in an honest
execution of the protocol. One of the results of this paper is the
generalization of Chailloux and Kerenidis' optimal Strong CF protocol
to cover the imbalanced case as well. Similalry, also in the weak
case, optimal protocols exist for every degree of imbalance \cite{Kerenidis},
thereby generalizing Mochon's aforementioned result.

Somewhat surprisingly, CF has yet to be generalized to many outcomes.
Indeed, even in classical settings this problem is nontrivial, as
a dishonest party can no longer force with certainty all of the outcomes.
In this paper we shall focus on two problems. In the first, which
we term weak dice rolling (DR), $N>2$ remote distrustful parties
must agree on a number between $1$ and $N$ with party $i$ preferring
the $i$-th outcome. In the second, which we term strong DR, $M$
remote distrustful parties must agree on a number between $1$ and
$N$ without any party being aware of any other's preference.\\

The paper is organized as follows. In section II we prove that, using
quantum resources, weak $N$-sided dice rolling with arbitrarily small
bias is possible for any value of $N$. This result stands in marked
contrast to the classical case, where, under certain conditions, an
honest party always loses. Furthermore, to gain insight as to what
biases are achievable with a minimal number of rounds of communication,
we present a six-round weak three-sided DR protocol, which incorporates
a three-round weak imbalanced CF protocol that generalizes the results
of Spekkens and Rudolph to the imbalanced case. In section III we
generalize Kitaev's bound to any number of parties, $M$, and outcomes,
$N$, and present a family of two-party protocols that saturate it
for any value of $N$. We then make use of this family to extend this
result to $2m$-party $n^{m}$-sided DR any for value of $m$ and
$n$. In the process, we generalize Chailloux and Kerenidis' optimal
strong CF protocol to cover the imbalanced case. Finally, we analyze
a family of three-round two-party strong $N$-sided DR protocols for
any value of $N$.

\section{Weak dice rolling with arbitrarily small bias}

The purpose of weak CF is to decide between two parties. Hence, its
natural multi-outcome generalization is the problem of deciding between
$N>2$ parties. As opposed to weak CF, in weak $N>2$-sided DR there
are many different cheating scenarios, as any number of parties $n<N$
may be dishonest. We shall be interested in the $N$ {}``worst case''
scenarios where all but one of the parties are dishonest and, moreover,
are acting in unison. That is, the dishonest parties share classical
and quantum communication channels and have a joint strategy. In addition,
we shall require that the protocol be {}``fair'' in the sense that
the honest party's maximum losing probability be the same in each
of these $N$ scenarios. Of course, the security of the protocol can
be evaluated with respect to any other cheating scenario, but as we
shall consider only fair protocols, the security of any cheating scenarios
is never poorer than that provided by the afore-mentioned $N$ worst
case scenarios.\\

We begin by observing that in CF the bias has a complementary definition.
Concentrating on weak CF, we could just as well define it as \begin{equation}
\bar{\epsilon}_{i}\hat{=}\bar{P}_{i}^{*}-1/2\,,\qquad i=A,\, B\end{equation}
 where $\bar{P}_{i}^{*}=P_{j\neq i}^{*}$ is the maximum probability
that party $i$ loses. According to this definition the bias tells
us to what extent party $j\neq i$ can increase other party $i$'s
chances of losing beyond one half. In the case of $N$ parties, the
bias $\bar{\epsilon}_{i}$ then tells us to what extent the $N-1$
dishonest parties can increase party $i$'s chances of losing beyond
$1-1/N$, rather than to what extent a sole dishonest party can increase
its chances of winning beyond $1/N$. We shall always use this redefinition
of the bias when considering weak DR. The computation of biases in
weak DR is therefore equivalent to the computation of biases in a
weak imbalanced coin flipping protocol.

\subsection{Weak dice rolling with arbitrarily small bias}

We shall now prove that quantum weak $N$-sided DR with arbitrarily
small bias is possible for any $N$. The proof is by construction.
Consider the following $N$-party protocol. Each party is uniquely
identified according to a number from $1$ to $N$. The protocol consists
of $N-1$ stages. In stage one parties $1$ and 2 {}``weakly flip''
a balanced quantum coin. The winner and party $3$ then weakly flip
an imbalanced quantum coin in stage two, where if both parties are
honest $3$'s winning probability equals $1/3$. And so on, the rule
being that in stage $n\geq2$ the winner of stage $n-1$ and party
$n$ {}``weakly flip'' an imbalanced quantum coin, where if both
parties are honest, $n$'s winning probability equals $1/\left(n+1\right)$.
Thus, if all parties are honest each has the same overall winning
probability of $1/N$. Using Mochon's formalism \cite{Mochon}, Chailloux
and Kerenidis have recently proved that weak imbalanced coin flipping
with arbitrarily small bias is possible \cite{Kerenidis}. It follows
that in the limit where each of the weak imbalanced coin flipping
protocols, used to implement our DR protocol, admits a vanishing bias
(and $N$ is finite), any honest party's winning probability tends
to $1/N$; for a formal proof see appendix A. Moreover, since we have
considered the worst case cheating scenario, this result holds for
any other cheating scenario.

The above result stands in stark contrast to the classical case where
if the number of honest parties is not strictly greater than $N/2$,
then the dishonest parties can force any outcome they desire. To see
why this is so, let us consider a classical $N$-sided dice rolling
protocol and partition the parties into two groups of $m\le\left\lceil N/2\right\rceil $
and $n=N-m$ parties. If both groups are honest, the probability that
a party in the first (second) group wins is $m/N$ ($1-m/N$). Therefore,
any weak DR protocol can serve as a weak imbalanced CF protocol. Suppose
now that all of the parties in the second group are dishonest, and
are nevertheless unable to force with certainty the outcome they choose.
Clearly, this would still be the case even if they were the smaller
group, i.e. $n<m$ ($m>\left\lceil N/2\right\rceil $), and we get
a contradiction, since in classical weak imbalanced CF (as in the
balanced case) at least one of the parties is always able to force
whichever outcome they desire \cite{Keyl}.

\subsection{A six-round weak three-sided dice rolling protocol}

Apart from the inherent limitations on the security of a multi-party
quantum cryptographic protocol, it is interesting, both from a theoretical
and a practical viewpoint, to determine what degree of security is
afforded using the least amount of communication. In this section
we introduce a six-round three-sided dice rolling protocol following
the general construction presented in the second section. This construction
gives rise to different biases dependent on the biases of the weak
imbalanced CF protocol employed in each of its stages. Three-round
weak imbalanced CF protocols to date have never been analyzed. In
the next subsection we carry out just such an analysis, which is then
used in the subsequent section obtain definite results for the DR
protocol.

\subsubsection*{A three-round weak imbalanced coin flipping protocol}

We introduce a three-round weak imbalanced coin flipping protocol
based on quantum gambling. It is constructed such that if both parties
are honest Alice's winning probability equals $1-p$. Interestingly,
it turns out that this protocol coincides with the generalization
of Spekkens and Rudolph's work to the imbalanced case. The protocol
consists of three rounds:
\begin{itemize}
\item Alice prepares a superposition of two qubits \begin{equation}
\left|\psi_{0}\right\rangle =\sqrt{1-p-\eta}\left|\uparrow_{1}\downarrow_{2}\right\rangle +\sqrt{p+\eta}\left|\downarrow_{1}\uparrow_{2}\right\rangle ,\qquad0\leq\eta\leq1-p\,,\end{equation}

where the subscripts serve to distinguish between the first and second
qubit and will be omitted when the distinction is clear. She then
sends the second qubit to Bob.

\item Bob carries out a unitary transformation $U_{\eta}$ on the qubit
he received and another qubit (labelled by the subscript $3$) prepared
in the state $\left|\downarrow\right\rangle $ such that \begin{equation}
\left|\uparrow_{2}\downarrow_{3}\right\rangle \rightarrow U_{\eta}\left|\uparrow_{2}\downarrow_{3}\right\rangle =\sqrt{\frac{p}{p+\eta}}\left|\uparrow_{2}\downarrow_{3}\right\rangle +\sqrt{\frac{\eta}{p+\eta}}\left|\downarrow_{2}\uparrow_{3}\right\rangle \,,\end{equation}
 and \begin{equation}
\left|\downarrow_{2}\uparrow_{3}\right\rangle \rightarrow U_{\eta}\left|\downarrow_{2}\uparrow_{3}\right\rangle =\sqrt{\frac{\eta}{p+\eta}}\left|\uparrow_{2}\downarrow_{3}\right\rangle -\sqrt{\frac{p}{p+\eta}}\left|\downarrow_{2}\uparrow_{3}\right\rangle \,,\end{equation}
 with $U_{\eta}$ acting trivially on all other states. The resulting
state is then \begin{equation}
\left|\psi_{1}\right\rangle =U_{\eta}\left|\psi_{0}\right\rangle =\sqrt{1-p-\eta}\left|\uparrow_{1}\downarrow_{2}\downarrow_{3}\right\rangle +\sqrt{p}\left|\downarrow_{1}\uparrow_{2}\downarrow_{3}\right\rangle +\sqrt{\eta}\left|\downarrow_{1}\downarrow_{2}\uparrow_{3}\right\rangle \,.\end{equation}
 Following this, he checks whether the second and third qubits are
in the state $\left|\uparrow_{2}\downarrow_{3}\right\rangle $.
\item Bob wins if he finds the qubits in the state $\left|\uparrow_{2}\downarrow_{3}\right\rangle $.
Alice then checks whether the first qubit is in the state $\left|\downarrow\right\rangle $,
in which case Bob passes the test. If Bob does not find the qubits
in the state $\left|\uparrow_{2}\downarrow_{3}\right\rangle $, he
asks Alice for the first qubit and checks whether all three qubits
are in the state \begin{equation}
\left|\xi\right\rangle \hat{=}\sqrt{\frac{1-p-\eta}{1-p}}\left|\uparrow_{1}\downarrow_{2}\downarrow_{3}\right\rangle +\sqrt{\frac{\eta}{1-p}}\left|\downarrow_{1}\downarrow_{2}\uparrow_{3}\right\rangle \,,\end{equation}
 in which case she passes the test.
\end{itemize}
As proved in appendix B Alice's maximal winning probability is given
by \begin{equation}
P_{A}^{*}=\max_{\delta}\left(\sqrt{\frac{\left(1-p-\eta\right)\left(1-\delta\right)}{1-p}}+\sqrt{\frac{\eta^{2}\delta}{\left(1-p\right)\left(p+\eta\right)}}\right)^{2}\,,\qquad\delta\in\left[0,\,1\right]\end{equation}
 while Bob's maximal winning probability is given by \begin{equation}
P_{B}^{*}=p+\eta\,.\end{equation}
 In the balanced case a protocol is fair if $P_{A}^{*}=P_{B}^{*}$.
We can play with $\eta$ to make $P_{A}^{*}$ and $P_{B}^{*}$ minimal
under this constraint. It is easy to show that the minimum then obtains
for $\eta=\left(\sqrt{2}-1\right)/2$. It follows that $\epsilon_{A}=\epsilon_{B}=\left(\sqrt{2}-1\right)/2$
and $P_{A}^{\mathrm{*}}=P_{B}^{*}=1/\sqrt{2}$.

\subsubsection*{A six-round weak three-sided dice rolling protocol with a bias of
0.181}

The protocol consists of two three-round stages. In the first stage,
we have Alice and Bob weakly flip a balanced quantum coin. Following
this, in the second stage, the winner and Claire weakly flip an imbalanced
quantum coin, such that if both parties are honest Claire's winning
probability equals $1/3$. The protocol is considered fair if $\bar{P}_{A}^{*}=\bar{P}_{B}^{*}=\bar{P}_{C}^{*}$.
Due to the protocol's symmetry with respect to the interchange of
Alice and Bob there are only two nonequivalent worst case scenarios,
i.e. either only Alice is honest or only Claire is honest. Using the
quantum gambling based protocol an honest Alice has a maximum chance
of $1-1/\sqrt{2}$ of progressing to the second stage. Therefore,
Alice's maximum losing probability is given by\begin{equation}
\bar{P}_{A}^{\mathrm{*}}=\frac{1}{\sqrt{2}}+\left(1-\frac{1}{\sqrt{2}}\right)\bar{\Pi}_{2/3}^{*}\,,\end{equation}
 while an honest Claire's maximum losing probability is given by \begin{equation}
\bar{P}_{C}^{\mathrm{*}}=\bar{\Pi}_{1/3}^{*}\,,\end{equation}
 with $\bar{\Pi}_{1/3}^{*}$ ($\bar{\Pi}_{2/3}^{*}$) the maximum
losing probability of the party with a winning probability of $1/3$
($2/3$) when both parties are honest. Hence, we require that \begin{equation}
\bar{\Pi}_{1/3}^{*}=\frac{1}{\sqrt{2}}+\left(1-\frac{1}{\sqrt{2}}\right)\bar{\Pi}_{2/3}^{*}\,.\end{equation}
 If we use the WCF protocol of the previous section to implement the
second stage, then $\bar{\Pi}_{1/3}^{*}$ and $\bar{\Pi}_{2/3}^{*}$,
and hence the $\bar{P}_{i}^{*}$, will depend on $\eta$. We then
have to minimize the $\bar{P}_{i}^{*}$ with respect to $\eta$ under
the constraint that they are all equal, or what is the same thing,
minimize $\bar{\Pi}_{1/3}^{*}$ under the constraint eq. (13). However,
there are two possible implementations. Either $1-p=2/3$ and the
second stage begins with Alice preparing the state $\sqrt{2/3-\eta}\left|\uparrow_{1}\downarrow_{2}\right\rangle +\sqrt{1/3+\eta}\left|\downarrow_{1}\uparrow_{2}\right\rangle $,
or else $1-p=1/3$ and the second stage begins with Claire preparing
the state $\sqrt{1/3-\eta}\left|\uparrow_{1}\downarrow_{2}\right\rangle +\sqrt{2/3+\eta}\left|\downarrow_{1}\uparrow_{2}\right\rangle $.
In the first case we have to compute \begin{equation}
\min_{\eta}\max_{\delta}\frac{1}{2}\left(\sqrt{\left(2-3\eta\right)\left(1-\delta\right)}+\sqrt{\frac{9\eta^{2}\delta}{\left(1+3\eta\right)}}\right)^{2}\end{equation}
 under the constraint that \begin{equation}
\max_{\delta}\frac{1}{2}\left(\sqrt{\left(2-3\eta\right)\left(1-\delta\right)}+\sqrt{\frac{9\eta^{2}\delta}{\left(1+3\eta\right)}}\right)^{2}=\frac{1}{\sqrt{2}}+\left(1-\frac{1}{\sqrt{2}}\right)\left(\frac{1}{3}+\eta\right)\,,\end{equation}
 while in the second case we have to compute \begin{equation}
\min_{\eta}\left(\frac{2}{3}+\eta\right)\end{equation}
 under the constraint that \begin{equation}
\left(\frac{2}{3}+\eta\right)=\frac{1}{\sqrt{2}}+\left(1-\frac{1}{\sqrt{2}}\right)\max_{\delta}\left(\sqrt{\left(1-3\eta\right)\left(1-\delta\right)}+\sqrt{\frac{9\eta^{2}\delta}{\left(2+3\eta\right)}}\right)\,.\end{equation}
 The first of these yields the lower bias $\bar{\epsilon}_{A}=\bar{\epsilon}_{B}=\bar{\epsilon}_{C}\simeq0.181$
corresponding to $\bar{P}_{A}^{*}=\bar{P}_{B}^{*}=\bar{P}_{C}^{*}\simeq0.848$.
The second yields a bias of $0.199$.

\section{Optimal two-party strong dice rolling \& beyond}

In this section we consider the problem of $M\geq2$ remote distrustful
parties having to decide on a number between $1$ and $N\geq3$, without
any party being aware of any other's preference. We generalize Kitaev's
bound, eq. (2), to apply to this case as well, and present a protocol
that saturates it for $M=2m$ parties and $N=n^{m}$ outcomes for
any value of $m$ and $n$. In particular, this implies the possibility
of optimal two-party strong $N$-sided DR protocols for any value
of $N$. To this end we also introduce a protocol that saturates Kitaev's
bound for strong imbalanced CF, eq. (2).\\

It is straightforward to adapt the original proofs of Kitaev's bound
\cite{Kitaev,Multi-party} to cover more than two parties and outcomes.
Instead, however, we note that strong DR can always be used to implement
strong imbalanced CF. In particular, let us consider an $M>2$-party
strong $N$-sided DR protocol. The probability for each of the outcomes
in an honest execution is $P_{i}=1/N$. Suppose that we take the first
$N-1$ outcomes (last outcome) to represent $0$ ($1$) in an $M$-party
strong imbalanced CF protocol, such that there is a $\left(N-1\right)/N$
probability of obtaining $0$ in an honest execution. Kitaev's bound
can be generalized to cover this case as well and reads $\left.\bar{P}_{A}^{\left(1\right)}\right.^{*}\cdot\left.\bar{P}_{B}^{\left(1\right)}\right.^{*}\cdot.\,.\,.\cdot\left.\bar{P}_{M}^{\left(1\right)}\right.^{*}\geq1/N$
\cite{Multi-party}, where now $\left.\bar{P}_{j}^{\left(1\right)}\right.^{*}$
gives the probability for the outcome $1$ when all parties but the
$j$-th are dishonest and acting in unison to force the outcome $1$.
It follows that this bound should apply to $M$-party strong $N$-sided
DR as well (for otherwise we get a contradiction). That is, \begin{equation}
\left.\bar{P}_{A}^{\left(i\right)}\right.^{*}\cdot\left.\bar{P}_{B}^{\left(i\right)}\right.^{*}\cdot.\,.\,.\cdot\left.\bar{P}_{M}^{\left(i\right)}\right.^{*}\geq\frac{1}{N}\,,\qquad i\in\left\{ 1,\,\dots,\, M\right\} \,.\end{equation}
In the case of a protocol which is symmetric in the biases, i.e. for
any value of $i$, $j$, $k$, and $l$ $\left.\bar{P}_{i}^{\left(j\right)}\right.^{*}=\left.\bar{P}_{k}^{\left(l\right)}\right.^{*}$,
we then have that \begin{equation}
q\geq\left(\frac{1}{N}\right)^{1/M}\,,\end{equation}
where $q$ now denotes the maximal probability of any of the $N-1$
parties to bias the result to any of the outcomes.

\subsection{Optimal strong imbalanced coin flipping}

To prove that the above bound can be saturated, we assume the existence
of a strong imbalanced CF protocol saturating Kitaev's bound, eq.
(2). Hence, we shall begin by presenting such a protocol, based on
Chailloux and Kerenidis' optimal strong CF protocol:
\begin{itemize}
\item Alice flips an imbalanced coin, such that $0$ obtains with a probability
$q$ and $1$ obtains with a probability $1-q$, and sends the outcome
$o$ to Bob.
\item If $o=0$ Alice and Bob carry out an optimal imbalanced weak CF protocol,
where if both parties are honest Alice wins with probability $z_{0}$
and Bob wins with probability $1-z_{0}$. If $o=1$ Alice and Bob
carry out an optimal imbalanced weak CF protocol, where if both parties
are honest Alice wins with probability $z_{1}$ and Bob wins with
probability $1-z_{1}$.
\item If Alice wins the (weak) coin flip, the outcome of the (strong CF)
protocol is $o$.
\item If Bob wins the coin flip, then he weakly flips an imbalanced coin,
whose degree of imbalance is dependent on $o$. When $o=0$, Bob flips
an imbalanced coin such that its outcome is equal to $0$ ($1$) with
probability $p_{0}$ ($1-p_{0}$). When $o=1$, Bob flips an imbalanced
such that its outcome equals $1$ ($0)$ with probability $p_{1}$
($1-p_{1})$. The outcome of this last coin flip is the outcome of
the protocol.
\end{itemize}
Denoting by $P_{i}$ the probability of the outcome $i$ when both
parties are honest, we have that

\begin{equation}
P_{0}=q\left(z_{0}+\left(1-z_{0}\right)p_{0}\right)+\left(1-q\right)\left(1-z_{1}\right)\left(1-p_{1}\right)\,,\end{equation}
 and $P_{1}=1-P_{0}$. This protocol differs from Chailloux and Kerenidis'
protocol in in that in the first round Alice performs an imbalanced
coin flip, rather than a balanced one, and dependent on its outcome,
she and Bob carry out different weak imbalanced CF protocols. In addition,
if Bob wins then dependent on the value of $o$ he flips one of two
different coins. Thus, instead of two free parameters we now have
five. It is this extra freedom that allows the generalization to any
degree of imbalance.

To obtain the biases, suppose that a dishonest Alice tries to bias
the outcome to $0$. There are two ways that this can be achieved,
either by announcing that she has obtained $o=0$ or by announcing
that she has obtained $o=1$. In the first case, her maximal probability
of success equals

\begin{equation}
\left.P_{A}^{\left(0\right)}\right.^{*}=z_{0}+\epsilon_{0}+\left(1-z_{0}-\epsilon_{0}\right)p_{0}\,,\end{equation}
 where $\epsilon_{0}\ll1$ is the bias of the weak imbalanced CF that
is carried out when Alice inputs $0$, while in the second case her
maximal probability of success equals

\begin{equation}
\left.Q_{A}^{\left(0\right)}\right.^{*}=1-p_{1}\,.\end{equation}
 Similarly, if Alice tries to bias the outcome to $1$, her maximal
probabilities of success equals \begin{equation}
\left.P_{A}^{\left(1\right)}\right.^{*}=z_{1}+\epsilon_{1}+\left(1-z_{1}-\epsilon_{1}\right)p_{1}\,,\end{equation}
 where $\epsilon_{1}\ll1$ is the bias of the weak imbalanced coin
flip. which is performed whenever Alice inputs $1$, and \begin{equation}
\left.Q_{A}^{\left(1\right)}\right.^{*}=1-p_{0}\,.\end{equation}
 Suppose now that a dishonest Bob tries to bias the outcome to $0$.
Given that in the first stage Alice outputs $0$ ($1)$ probability
$q$ ($1-q$), Bob's maximal probability of success is given by

\begin{equation}
\left.P_{B}^{\left(0\right)}\right.^{*}=q+\left(1-q\right)\left(1-z_{1}+\epsilon_{1}\right)\,,\end{equation}
 while if he tries to bias the outcome to $1$, his maximal probability
of success is given by

\begin{equation}
\left.P_{B}^{\left(1\right)}\right.^{*}=1-q+q\left(1-z_{0}+\epsilon_{0}\right)\,.\end{equation}
 For ideal weak imbalanced CF (i.e. $\epsilon_{0}=\epsilon_{1}=0$)
this construction allows for Kitaev's bound to be exactly attained.
This can be seen by imposing the following four constraints \begin{equation}
\left.P_{A}^{\left(i\right)}\right.^{*}=\left.Q_{A}^{\left(i\right)}\right.^{*}\,,\qquad i=\left\{ 0,\,1\right\} \end{equation}
 \begin{equation}
\left.P_{A}^{\left(i\right)}\right.^{*}=\left.P_{B}^{\left(i\right)}\right.^{*}\,,\qquad i\in\left\{ 0,\,1\right\} \end{equation}
 Solving these equations together with eq. (20) we get \begin{equation}
q=\frac{1}{2}\left(1+\sqrt{P_{0}}-\sqrt{P_{1}}\right)\,,\end{equation}

\begin{equation}
p_{0}=1-\sqrt{P_{1}}\,,\qquad p_{1}=1-\sqrt{P_{0}}\,,\end{equation}

\begin{equation}
z_{1}=1+\frac{\sqrt{P_{0}}-1}{\sqrt{P_{1}}}\,,\qquad z_{2}=1+\frac{\sqrt{P_{1}}-1}{\sqrt{P_{0}}}\,.\end{equation}
 Note that for $P_{i}\in\left[0,\,1\right]$ $q$, $z_{i}$, and $p_{i}$
also in the required range of values, i.e. $\left[0,\,1\right]$.
Substituting back into eqs. (21) to (26) we get

\begin{equation}
\left.P_{A}^{\left(0\right)}\right.^{*}=\left.P_{B}^{\left(0\right)}\right.^{*}=\sqrt{P_{0}}\,,\qquad\left.P_{A}^{\left(1\right)}\right.^{*}=\left.P_{B}^{\left(1\right)}\right.^{*}=\sqrt{P_{1}}\,.\end{equation}
 Returning to the non-ideal case, using for $q$, the $z_{i}$, and
the $p_{i}$ the values just obtained, from eqs. (21) to (26) we have
that

\begin{equation}
\left.P_{A}^{\left(0\right)}\right.^{*}=\left.Q_{A}^{\left(0\right)}\right.^{*}+\epsilon_{0}\sqrt{P_{1}}=\sqrt{P_{0}}+\epsilon_{0}\sqrt{P_{1}}\,,\qquad\left.P_{A}^{\left(1\right)}\right.^{*}=\left.Q_{A}^{\left(1\right)}\right.^{*}+\epsilon_{1}\sqrt{P_{0}}=\sqrt{P_{1}}+\epsilon_{1}\sqrt{P_{0}}\,,\end{equation}

\begin{equation}
\left.P_{B}^{\left(0\right)}\right.^{*}=\sqrt{P_{0}}+\frac{1}{2}\epsilon_{1}\left(1-\sqrt{P_{0}}+\sqrt{P_{1}}\right)\,,\qquad\left.P_{B}^{\left(1\right)}\right.^{*}=\sqrt{P_{1}}+\frac{1}{2}\epsilon_{0}\left(1+\sqrt{P_{0}}-\sqrt{P_{1}}\right)\,.\end{equation}
 (Note that we no longer require that the constraints eqs. (27) and
(28) be satisfied.) Since the $\epsilon_{i}$ can be made arbitrarily
small, it follows that the protocol saturates Kitaev's bound for any
degree of imbalance.

\subsection{Optimal two-party strong dice rolling}

Equipped with the above result we proceed to prove the possibility
of two-party strong $N$-sided DR saturating Kitaev's bound for any
value of $N$. Consider the following strong $N$-sided DR protocol.
In the first round the parties carry out a strong imbalanced CF protocol
such that there is a $\left\lceil N/2\right\rceil $ ($\left\lfloor N/2\right\rfloor $)
probability for the outcome $0$ ($1$). If the outcome of the coin
flip is $0$ ($1$), then they agree that the DR protocol's outcome
is (is not) going to lie between $1$ and $\left\lceil N/2\right\rceil $.
Suppose that the first coin flip results in $0$. Then in the second
round they {}``strongly'' flip another coin such that there is a
$\left\lceil \left\lceil N/2\right\rceil /2\right\rceil $ ($\left\lfloor \left\lceil N/2\right\rceil /2\right\rfloor $)
probability for the outcome $0$ ($1$). If the outcome is $0$ (1)
then they agree that the DR protocol's outcome is going to lie between
$1$ and $\left\lceil \left\lceil N/2\right\rceil /2\right\rceil $
($\left\lceil \left\lceil N/2\right\rceil /2\right\rceil +1$ and
$\left\lceil N/2\right\rceil $), and so on until they obtain a single
result (see Fig. 1). The probability of obtaining $1$ in an honest
execution equals \begin{equation}
\frac{\left\lceil N/2\right\rceil }{N}\cdot\frac{\left\lceil \left\lceil N/2\right\rceil /2\right\rceil }{\left\lceil N/2\right\rceil }\cdot\dots\cdot\frac{1}{\left\lceil \dots\left\lceil \left\lceil N/2\right\rceil /2\right\rceil \dots/2\right\rceil }=\frac{1}{N}\end{equation}
 It is straightforward to verify that this probability is true of
all other outcomes. Let us now consider a dishonest execution of the
protocol such that the biases of the underlying strong imbalanced
CF protocols are all equal to $\delta\ll1/\left\lceil \log_{2}N\right\rceil $.
The probability of obtaining the outcome $1$ is given by

\begin{equation}
\left(\sqrt{\frac{\left\lceil N/2\right\rceil }{N}}+\delta\right)\left(\sqrt{\frac{\left\lceil \left\lceil N/2\right\rceil /2\right\rceil }{\left\lceil N/2\right\rceil }}+\delta\right)\dots\left(\sqrt{\frac{1}{\left\lceil \dots\left\lceil \left\lceil N/2\right\rceil /2\right\rceil \dots/2\right\rceil }}+\delta\right)\simeq\frac{1}{\sqrt{N}}+c\left\lceil \log_{2}N\right\rceil \delta+O\left(\delta^{2}\right)\,,\end{equation}
 where $c\sim\sqrt{2/N}$. (The formal proof follows along the same
lines as that given in the appendix for weak DR, and so is omitted.)
Similar expressions obtain for the probabilities of all other outcomes.
Hence, we have shown that the this construction saturates the generalization
of Kitaev's bound, eq. (18), for $M=2$ and any $N$.%
\begin{figure}
\center{ \includegraphics[scale=0.45]{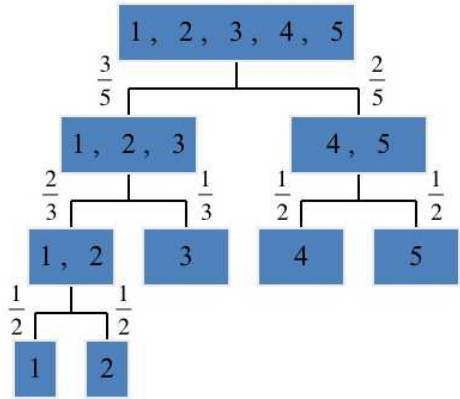}

\caption{Two-party strong five-sided DR protocol saturating Kitaev's bound.
The digits inside the boxes denote the possible outcomes. Each branching
represents a strong imbalanced coin flip. The fractions beside each
branch give the probability for the outcomes within the box below
conditional on the outcomes in the box above. Thus, for the leftmost
branch we have that the probability for the outcome equals $1/2\cdot2/3\cdot3/5=1/5$,
etc.}

}
\end{figure}

\subsection{A family of optimal multi-party strong dice rolling protocols}

The above construction readily allows for the introduction of a family
of $2m$-party strong $n^{m}$-sided DR protocols saturating the generalization
of Kitaev's bound, eq. (18). The idea is to sequentially have distinct
pairs of parties strongly roll a dice to eliminate some of the outcomes,
until a single outcome is obtained. Thus, in the case of four parties
and nine outcomes, in the first stage the first and second parties
strongly roll a three-sided dice. If its outcome is $1$, outcomes
$4$ to $9$ are eliminated, while if its outcome is $2$, outcomes
$1$ to $3$ and $7$ to $9$ are eliminated, etc. Suppose, for example,
that outcomes $1$ to $6$ are eliminated. Then in the second stage
parties three and four strongly roll a three-sided dice, where if
its outcome is $1$, then the final outcome of the protocol is $7$,
while if its outcome is $2$, then the final outcome is $8$, etc.
In general, for $2m$-parties and an $n^{m}$-sided dice, the protocol
consists of $n$ stages. In each stage a different pair of parties
strongly rolls an $n$-sided dice. As there are a total of 2$m$ parties,
each party strongly participates in a dice roll once. In order to
force the outcome they desire, the $2m-1$ dishonest parties must
bias the result of the dice roll in which the honest party participates,
and since at any stage there is only a single outcome that can lead
to the desired outcome, the dishonest parties can maximally bias the
outcome with a probability of $\left(1/n\right)^{1/2}+\bar{\epsilon}$,
which saturates the generalization of Kitaev's bound, eq. (18).

It is not straightforward to generalize this scheme to any number
of parties and outcomes. The problem is that we have introduced an
ordering, which dependent on it, may in general render the protocol
asymmetric in the biases, or even trivial by allowing the dishonest
parties to force the outcome that they desire. This can be fixed by
making use of optimal weak DR to decide the ordering. Unfortunately,
this comes at the expense of optimality, i.e. eq. (18) is no longer
saturated. Nevertheless, protocols incorporating optimal weak and
strong DR protocols may give rise to biases remarkably close to the
inherent bounds. As an example, consider the following three-party
strong three-sided DR protocol. In the first round Alice, Bob and
Clare weakly roll a three-sided dice. The winner then randomly selects
a number $a\in\left\{ 1,\,3\right\} $ and informs the two losers
of his/her choice. The two losers then strongly flip a coin. Denote
its outcome by $b\in\left\{ 0,\,1\right\} $. The outcome of the protocol
is $\left(a+b\right)\,\mathrm{mod}\,3$. It is easy to verify that
the maximal probability of any two parties to successfully bias to
any of the outcomes approximately equals $0.69363$, while from eq.
(19) $q=\left(1/3\right)^{1/3}\simeq0.69336$. That is, a difference
of $0.027\,\%$. Similarly, to the optimal $2m$-party $n^{m}$-sided
DR protocols described above, this protocol can be generalized to
a family of $3n$-party $3^{n}$-sided DR protocol, with each giving
rise to the same bias.\\

To complete the discussion we should mention that strong DR is nontrivial
also in classical settings. Indeed, the classical biases for two-party
$N$-sided DR are constrained by the following set of inequalities
\cite{Kitaev} \begin{equation}
\left(1-\left.\bar{P}_{A}^{\left(i\right)}\right.^{*}\right)\left(1-\left.\bar{P}_{B}^{\left(j\right)}\right.^{*}\right)\leq\frac{N-2}{N}+\frac{1}{N}\delta_{i,\, j}\,,\qquad i,\, j\in\left\{ 1,\,\dots,\, N\right\} \,.\end{equation}
 It is straightforward to verify that these inequalities are {}``weaker''
than the corresponding Kitaev bound, and hence allow for higher biases.

\subsection{A family of three-round two-party strong dice rolling protocols}

In this subsection we introduce a family of three-round strong DR
protocols, which generalizes Colbeck's entanglement-based strong CF
protocol \cite{Colbeck} to any number of outcomes. Suppose Alice
and Bob want to strongly roll an $N$-sided dice using a minimal number
of rounds of communication. Then they may proceed as follows. Alice
prepares a pair of systems in the state $\left|\psi_{N}\right\rangle \otimes\left|\psi_{N}\right\rangle $,
where $\left|\psi_{N}\right\rangle =\frac{1}{\sqrt{N}}\sum_{i=1}^{N}\left|i\right\rangle \otimes\left|i\right\rangle $,
and sends the second half of each system to Bob. Bob randomly selects
one of the systems to serve as the dice and informs Alice of his selection.
Alice and Bob then measure their half of the selected system in the
Schmidt basis. The outcome of this measurement is the outcome of the
dice roll. Finally, Alice sends Bob her half of system that was not
selected, and he verifies it was indeed prepared in the state $\left|\psi_{N}\right\rangle $.

Following a similar argument to Colbeck's, Alice's and Bob's maximal
probabilities of biasing to any of the outcomes are given by

\begin{equation}
\left.P_{A}^{\left(i\right)}\right.^{*}=\frac{N+1}{2N}\,,\qquad\left.P_{B}^{\left(i\right)}\right.^{*}=\frac{2N-1}{N^{2}}\,,\qquad\left\{ i=1,\,\dots,\, N\right\} \end{equation}
 Thus, for $N=3$ $\epsilon_{A}^{\left(i\right)}=\left.P_{A}^{\left(i\right)}\right.^{*}-1/3=1/3$
and $\epsilon_{B}^{\left(i\right)}=\left.P_{B}^{\left(i\right)}\right.^{*}-1/3=2/9$.
Interestingly, in the limit where $N\rightarrow\infty$, $\left.P_{A}^{\left(i\right)}\right.^{*}\rightarrow1/2$
, $\left.P_{B}^{\left(i\right)}\right.^{*}\rightarrow2/N$, so that
$\left.P_{A}^{\left(i\right)}\right.^{*}\cdot\left.P_{B}^{\left(i\right)}\right.^{*}\rightarrow1/N$.
Hence, in this limit Kitaev's bound is nontrivially saturated in a
finite number of rounds, albeit at a cost of a high asymmetry of the
biases.

\section{Conclusions}

We have defined a novel mutli-outcome generalization of quantum CF,
which we have termed quantum DR. We have analyzed both its weak and
strong variants. Specifically, we proved by construction that in quantum
settings (i) weak $N$-sided dice rolling -- the problem of $N$ remote
distrustful parties having to decide on a number between $1$ and
$N$ with party $i$ preferring the $i$-th outcome -- admits an arbitrarily
small bias for any value of $N$, and (ii) two-party strong $N$-sided
dice rolling -- the problem of two remote distrustful parties having
to decide on a number between $1$ and $N$ without any party being
aware of the other's preference -- saturates the corresponding generalization
of Kitaev's bound for any value of $N$. In addition, we also made
use of this last result to introduce a family of optimal $2m$-party
strong $n^{m}$-sided DR protocols for any value of $m$ and $n$.
The question of whether this is also possible in the general case
of any number of parties and outcomes remains open for now.
\begin{acknowledgments}
We thank Lev Vaidman and Oded Regev for useful comments. N. Aharon
acknowledges the support of the Wolfson Foundation. J. Silman acknowledges
the support of the the Israeli Science Foundation.
\end{acknowledgments}
\appendix

\section{}

For any weak DR protocol, based on weak imbalanced CF according to
the scheme presented in section II, party $n$'s maximum chance of
losing is given by \begin{eqnarray}
\bar{P}_{n}^{*} & = & \frac{N-1}{N}+\bar{\epsilon}_{n}\\
 & = & \mbox{\ensuremath{\bar{\Pi}}}_{n-1}^{*}+\sum_{k=n}^{N-1}\mbox{\ensuremath{\bar{\Pi}}}_{k}^{*}\prod_{j=0}^{k-n}\left(1-\bar{\Pi}_{n-1+j}^{*}\right)\nonumber \\
 & = & \frac{n-1}{n}+\bar{\delta}_{n-1}+\sum_{k=n}^{N-1}\left(\frac{1}{k}+\bar{\delta}_{k}\right)\left(\frac{1}{n}-\bar{\delta}_{n-1}\right)\prod_{j=1}^{k-n-1}\left(\frac{n+j}{n+j+1}-\bar{\delta}_{n-1+j}\right)\,,\nonumber \end{eqnarray}
 where $\bar{\Pi}_{k}^{*}$ is party $n$'s maximum chance of losing
stage $k$ conditional on having made it to that round and $\bar{\delta}_{k}$
the corresponding bias. If we now let $\bar{\delta}_{\mathrm{max}}^{\left(n\right)}\hat{=}\max_{k}\bar{\delta}_{k}$
and $\bar{\delta}_{\mathrm{min}}^{\left(n\right)}\hat{=}\min_{k}\bar{\delta}_{k}$
($k=n-1,\,\dots,\, N-1$), then \begin{eqnarray}
\bar{\epsilon}_{n} & \leq & \bar{\delta}_{\mathrm{max}}^{\left(n\right)}+\bar{\delta}_{+}^{\left(n\right)}\sum_{k=n}^{N-1}\left(\frac{1}{n}-\bar{\delta}_{\mathrm{min}}^{\left(n\right)}\right)\prod_{j=1}^{k-n-1}\left(\frac{n+j}{n+j+1}-\bar{\delta}_{\mathrm{min}}^{\left(n\right)}\right)-\bar{\delta}_{\mathrm{min}}^{\left(n\right)}\sum_{k=n}^{N-1}\left(\frac{1}{k}+\bar{\delta}_{\mathrm{max}}^{\left(n\right)}\right)\prod_{j=1}^{k-n-1}\left(\frac{n+j}{n+j+1}-\bar{\delta}_{\mathrm{min}}^{\left(n\right)}\right)\nonumber \\
 &  & -\bar{\delta}_{\mathrm{min}}^{\left(n\right)}\sum_{k=n}^{N-1}\left(\frac{1}{k}+\bar{\delta}_{\mathrm{max}}^{\left(n\right)}\right)\left(\frac{1}{n}-\bar{\delta}_{\mathrm{min}}^{\left(n\right)}\right)\sum_{m=1}^{k-n-1}\prod_{j\neq m}\left(\frac{n+j}{n+j+1}-\bar{\delta}_{\mathrm{min}}^{\left(n\right)}\right)\nonumber \\
 & < & \bar{\delta}_{\mathrm{max}}^{\left(n\right)}+\bar{\delta}_{\mathrm{max}}^{\left(n\right)}\sum_{k=n}^{N-1}\frac{1}{n}\prod_{j=1}^{k-n-1}\frac{n+j}{n+j+1}\\
 & < & N\bar{\delta}_{\mathrm{max}}^{\left(n\right)}\nonumber \end{eqnarray}
 Hence, if each of the weak imbalanced CF protocols, used to implement
the DR protocol, are such that $\bar{\delta}_{\mathrm{max}}^{\left(n\right)}\ll1/N$
for any $n$, an honest party's winning probability tends to $1/N$.

\section{}

\subsubsection*{Alice's maximal bias}

Most generally Alice can prepare any state of the form \begin{equation}
\left|\psi_{0}'\right\rangle =\sum_{i,\, j=\uparrow,\,\downarrow}\alpha_{ij}\left|ij\right\rangle \otimes\left|\Phi_{ij}\right\rangle \,,\end{equation}
 where the $\left|\Phi_{ij}\right\rangle $ are states of some ancillary
system at her possession. After Bob applies $U_{\eta}$ the resulting
composite state is given by \begin{eqnarray}
\left|\psi_{1}'\right\rangle  & = & U_{\eta}\left|\psi_{0}'\right\rangle \otimes\left|\downarrow\right\rangle \\
 & = & \alpha_{\uparrow\uparrow}\left(\sqrt{\frac{p}{p+\eta}}\left|\uparrow\uparrow\downarrow\right\rangle +\sqrt{\frac{\eta}{p+\eta}}\left|\uparrow\downarrow\uparrow\right\rangle \right)\otimes\left|\Phi_{\uparrow\uparrow}\right\rangle +\alpha_{\uparrow\downarrow}\left|\uparrow\downarrow\downarrow\right\rangle \otimes\left|\Phi_{\uparrow\downarrow}\right\rangle \nonumber \\
 &  & +\alpha_{\downarrow\uparrow}\left(\sqrt{\frac{p}{p+\eta}}\left|\downarrow\uparrow\downarrow\right\rangle +\sqrt{\frac{\eta}{p+\eta}}\left|\downarrow\downarrow\uparrow\right\rangle \right)\otimes\left|\Phi_{\downarrow\uparrow}\right\rangle +\alpha_{\downarrow\downarrow}\left|\downarrow\downarrow\downarrow\right\rangle \otimes\left|\Phi_{\downarrow\downarrow}\right\rangle \,.\nonumber \end{eqnarray}
 The probability that Bob does not find find the second and third
qubits in the state $\left|\uparrow_{2}\downarrow_{3}\right\rangle $
is \begin{equation}
\bar{P}_{\uparrow\downarrow}=1-P_{\uparrow\downarrow}=1-\frac{\left|\alpha_{\uparrow\uparrow}\right|^{2}p+\left|\alpha_{\downarrow\uparrow}\right|^{2}p}{p+\eta}\,,\end{equation}
 and the resulting composite state then is \begin{eqnarray}
\left|\psi_{2}'\right\rangle  & = & \mathcal{N}\left(\alpha_{\uparrow\uparrow}\sqrt{\frac{\eta}{p+\eta}}\left|\uparrow\downarrow\uparrow\right\rangle \otimes\left|\Phi_{\uparrow\uparrow}\right\rangle +\alpha_{\uparrow\downarrow}\left|\uparrow\downarrow\downarrow\right\rangle \otimes\left|\Phi_{\uparrow\downarrow}\right\rangle \right.\nonumber \\
 &  & \left.+\alpha_{\downarrow\uparrow}\sqrt{\frac{\eta}{p+\eta}}\left|\downarrow\downarrow\uparrow\right\rangle \otimes\left|\Phi_{\downarrow\uparrow}\right\rangle +\alpha_{\downarrow\downarrow}\left|\downarrow\downarrow\downarrow\right\rangle \otimes\left|\Phi_{\downarrow\downarrow}\right\rangle \right)\,,\end{eqnarray}
 where $\mathcal{N}$, the normalization, is \begin{equation}
\frac{1}{\mathcal{N}^{2}}=1-\frac{p}{p+\eta}\left(\left|\alpha_{\uparrow\uparrow}\right|^{2}+\left|\alpha_{\downarrow\uparrow}\right|^{2}\right)\,.\end{equation}
 The probability that Alice passes the test is therefore given by
\begin{equation}
P_{\mathrm{test}}=\left\Vert \left\langle \xi\mid\psi_{2}'\right\rangle \right\Vert ^{2}=\mathcal{N}^{2}\left\Vert \alpha_{\uparrow\downarrow}\sqrt{\frac{1-p-\eta}{1-p}}\left|\Phi_{\uparrow\downarrow}\right\rangle +\alpha_{\downarrow\uparrow}\sqrt{\frac{\eta^{2}}{\left(1-p\right)\left(p+\eta\right)}}\left|\Phi_{\downarrow\uparrow}\right\rangle \right\Vert ^{2}\,.\end{equation}
 The maximum obtains for $\left|\Phi_{\uparrow\downarrow}\right\rangle =\left|\Phi_{\downarrow\uparrow}\right\rangle $.
This choice of the ancillary states does not affect the maximum of
$\bar{P}_{\uparrow\downarrow}$. Hence, Alice obtains no advantage
by using ancillary systems and we can do away with them. Alice's maximum
cheating probability is then \begin{equation}
P_{A}^{*}=\max_{\alpha_{ij}}\bar{P}_{\uparrow\downarrow}\cdot P_{\mathrm{test}}\,,\end{equation}
 where now \begin{equation}
\bar{P}_{\uparrow\downarrow}\cdot P_{\mathrm{test}}=\left|\alpha_{\uparrow\downarrow}\sqrt{\frac{1-p-\eta}{1-p}}+\alpha_{\downarrow\uparrow}\sqrt{\frac{\eta^{2}}{\left(1-p\right)\left(p+\eta\right)}}\right|^{2}\end{equation}
 ($\bar{P}_{\uparrow\downarrow}=1/\mathcal{N}^{2}$). Clearly, this
expression is maximum when $\alpha_{\uparrow\uparrow}=\alpha_{\downarrow\downarrow}=0$.
Therefore, to maximize her chance of successfully cheating Alice will
prepare a state of the form \begin{equation}
\left|\psi_{0}'\right\rangle =\sqrt{1-\delta}\left|\uparrow_{1}\downarrow_{2}\right\rangle +\sqrt{\delta}\left|\downarrow_{1}\uparrow_{2}\right\rangle \,,\end{equation}
 where with no loss of generality we have set $\alpha_{\uparrow\downarrow}=\sqrt{1-\delta}$
and $\alpha_{\downarrow\uparrow}=\sqrt{\delta}$. So that \begin{equation}
P_{A}^{*}=\max_{\delta}\left(\sqrt{\frac{\left(1-p-\eta\right)\left(1-\delta\right)}{1-p}}+\sqrt{\frac{\eta^{2}\delta}{\left(1-p\right)\left(p+\eta\right)}}\right)^{2}\,.\end{equation}

\subsubsection*{Bob's maximal bias}

Bob wins and passes the test whenever Alice does not find the first
qubit in the state $\left|\uparrow\right\rangle $. The probability
for this is just $p+\eta$. This gives an upper bound on Bob's maximal
cheating probability, which is reached if Bob always announces that
he has won. That is, \begin{equation}
P_{B}^{*}=p+\eta\,.\end{equation}


\begin{thebibliography}{18}
\bibitem{footnote}Prior to Mochon's work \cite{Mochon}, there were
serious doubts as to the possibility of WCF with arbitrarily small
bias. Quantum gambling was introduced as an example of a quantum cryptographic
task, impossible in classical settings, which is nevertheless unconditionally
secure in the sense that it does away with the notion of cheating;
both the casino and the player do not have to follow any {}``honest''
protocol, but rather are allowed to do anything to maximize their
gain.
\end{thebibliography}
\end{document}